\documentclass[aps,prl,twocolumn,superscriptaddress]{revtex4}

\usepackage{graphicx}
\usepackage[german,english]{babel}
\newcommand{\ang}{\ensuremath{\text{\AA}}}

\newcommand{\HHO}{\ensuremath{\text{H}_2\text{O}}}
\newcommand{\NOO}{\ensuremath{\text{N}\text{O}_2}}
\newcommand{\NOOd}{\ensuremath{\text{N}_2\text{O}_4}}
\newcommand{\NHHH}{\ensuremath{\text{N}\text{H}_3}}

\begin{document}

\title{Molecular Doping of Graphene}

\author{T. O. Wehling}
\affiliation{I. Institute for Theoretical Physics, Hamburg University,
 Jungiusstra{\ss}e 9, D-20355 Hamburg, Germany}
\author{K. S. Novoselov}
\affiliation{School of Physics and Astronomy, University of Manchester, M13 9PL, Manchester, UK}
\author{S. V. Morozov}
\affiliation{Institute for Microelectronics Technology, 142432 Chernogolovka, Russia}
\author{E. E. Vdovin}
\affiliation{Institute for Microelectronics Technology, 142432 Chernogolovka, Russia}
\author{M. I. Katsnelson}
\affiliation{Institute for Molecules and Materials, Radboud
University of Nijmegen, Toernooiveld 1, 6525 ED Nijmegen, The
Netherlands}
\author{A. K. Geim}
\affiliation{School of Physics and Astronomy, University of Manchester, M13 9PL, Manchester, UK}
\author{A. I. Lichtenstein}
\affiliation{I. Institute for Theoretical Physics, Hamburg University,
 Jungiusstra{\ss}e 9, D-20355 Hamburg, Germany}

%\thanks{}

\date{\today}

%\begin{abstract}
%\textbf{Graphene, a single layer of carbon atoms arranged in a two-dimensional (2D) honeycomb lattice, has recently been fabricated \cite{K.S.Novoselov07262005,Novoselov_science2004} and was demonstrated to exhibit remarkable physical properties. Electrons in graphene, obeying a linear dispersion relation, behave like massless relativistic particles, which results in a number of very peculiar electronic properties: from an anomalous quantum Hall effect to the absence of Anderson localization and exotic ``chiral'' tunnelling \cite{kostya2,kim,gus,per,cas,bilayer,FC,zitter1,carlo,ktsn}. The charge carrier concentration and mobility in graphene is determined by various impurities. Here we investigate, both experimentally and theoretically, the doping of graphene at the adsorption of {\NOO}. The existence of a magnetic moment for a single molecule is of crucial importance making this molecule a strong acceptor, in contrast with its diamagnetic dimer {\NOOd}. This explains recent results on the ``chemical sensor'' properties of graphene, in particular, the possibility to detect a single {\NOO} molecule \cite{schedin-gassensors}.}
%\end{abstract}

%\pacs{1234567}
%\keywords{}

\maketitle
\textbf{Graphene, a one-atom thick zero gap semiconductor \cite{Novoselov_science2004,K.S.Novoselov07262005}, has been attracting an increasing interest due to its remarkable physical properties ranging from an electron spectrum resembling relativistic dynamics \cite{kostya2,kim,gus,per,cas,bilayer,FC,zitter1,carlo,ktsn} to ballistic transport under ambient conditions \cite{Novoselov_science2004,K.S.Novoselov07262005,kostya2,kim}. The latter makes graphene a promising material for future electronics and the recently demonstrated possibility of chemical doping without significant change in mobility has improved graphene's prospects further \cite{schedin-gassensors}. However, to find optimal dopants and, more generally, to progress towards graphene-based electronics requires understanding the physical mechanism behind the chemical doping, which has been lacking so far.
Here, we present the first joint experimental and theoretical investigation of adsorbates on graphene. We elucidate a general relation between the doping strength and whether or not adsorbates have a magnetic moment: The paramagnetic single {\NOO} molecule is found to be a strong acceptor, whereas its diamagnetic dimer {\NOOd} causes only weak doping. This effect is related to the peculiar density of states of graphene, which provides an ideal situation for model studies of doping effects in semiconductors. Furthermore, we explain recent results on its ``chemical sensor'' properties, in particular, the possibility to detect a single {\NOO} molecule \cite{schedin-gassensors}.}

Controlling the type and the concentration of charge carriers is
at the heart of modern electronics: It is the ability of 
combining gate voltages and impurities for locally changing the
density of electrons or holes that allows for the variety of
nowadays available semiconductor based devices. 
However, the conventional Si-based electronics is expected to encounter fundamental limitations at the spatial scale below $10$\,nm, according to the semiconductor industry roadmap, and this calls for novel materials that might substitute or complement Si. Being only one atomic layer thick, graphene exhibits ballistic transport on a submicron scale and can be doped heavily --- either by gate voltages or molecular adsorbates --- without significant loss of mobility \cite{Novoselov_science2004}. In addition, later experiments \cite{schedin-gassensors} demonstrated its potential for solid state gas sensors and even the possibility of single molecule detection.

%For optimizing both, the novel electronic devices and the gas sensor applications, a microscopic understanding of the mechanisms behind the doping is crucial. 
We show, that in graphene, aside from the donor-acceptor distinction, there are in general two different classes of dopants --- paramagnetic and nonmagnetic. In contrast to ordinary semiconductors, the latter type of impurities act generally as rather weak dopants, whereas the paramagnetic impurities cause strong doping: Due to the linearly vanishing, electron-hole symmetric density of states (DOS) near the Dirac point of graphene, localised impurity states without spin polarisation are pinned to the centre of the pseudogap. Thus impurity states in graphene distinguish strongly from their counterparts in usual semiconductors, where the DOS in the valence and conduction bands are very different and impurity levels lie generally far away from the middle of the gap.
%In this sense, graphene provides a kind of model system for studying doping.

Impurity effects on the electronic structure of two-dimensional
systems with Dirac spectrum were investigated in detail in
connection with the problem of high-temperature superconductivity
in copper-oxide compounds, and STM visualization of the
order parameter around impurities is one of the most
straightforward evidences of the $d$-wave pairing in these systems
\cite{balatsky:373}. For the case of a strong enough impurity
potential, the formation of quasilocalized electron states in the
pseudogap around the Dirac point is expected \cite{balatsky:373},
whereas a weak potential will not lead to the formation of
quasilocalized states at all. This means that, in general, in this
situation one can hardly expect a strong doping effect which
requires existence of well-defined donor (or acceptor) levels
several tenths of electron volt away from the Fermi level. However,
if the impurity has a local magnetic moment its energy levels
split more or less symmetrically by the Hund exchange, of the order of
1 eV, which provides exactly the situation favorable for a strong
doping. To check this assumption we have chosen the {\NOO} system
forming both paramagnetic single molecules and diamagnetic dimers
{\NOOd}. Note that magnetism of impurity atoms in graphene is
interesting not only due to its effect on the electronic
properties but also itself. As was shown recently
\cite{Edwards_2006} ferromagnetism based on spin polarization of
$sp$-electrons in narrow impurity bands should be characterized by
much higher Curie temperatures than typical for conventional
magnetic semiconductors.

Doping and gas sensing effects in graphene related systems attracted a lot of research activity in the last years. Already the first
experiments with graphene showed the possibility of inducing charge carriers to this material by the adsorption of various gases including {\NHHH}, {\HHO} and {\NOO} \cite{Novoselov_science2004}. Hall effect measurements proved, that {\NHHH} induces electrons,
whereas the latter two types of adsorbates result in holes as
charge carriers. Furthermore, they showed that the detection of those gases at remarkably low concentrations
or for {\NOO} even in the extreme limit of single molecules is possible
\cite{schedin-gassensors}.

%Therefore, we present an \textit{ab-initio theory} and experiments revealing these mechanisms at the molecular level. The possibility of high temperature ferromagnetism in graphene is discussed as possible application.
%The proximate explanation of these steps as single molecules being detected raises the question of the microscopic mechanism beyond this effect, especially as it occurs solely for {\NOO} but not for CO, {\NHHH}, {\HHO}. This selectivity of the graphene decives manifests itself also in the delayed response of the sensors to the latter types of gases.

Carbon nanotubes (CNT) being rolled up graphene sheets exhibit
similar doping effects on gas exposure \cite{Kong:2000} and
stimulated first principles studies of these systems: DFT
calculations for {\NOO}, {\HHO} and {\NHHH} on nanotubes revealed
possible physisorbed geometries on nondefective CNTs and developed
a ``standard model'' to interpret this doping \cite{Peng2000,
chang2001, Zhao2002}: By considering of Mulliken or L\"owdin
charges of the adsorbed molecules, {\NOO} is found to accept 0.1
e$^-$ per molecule from the tube, whereas one {\NHHH} molecule is
predicted to donate between $0.03$ to $0.04$ e$^-$
\cite{chang2001, Zhao2002}. However, this ``standard model'' for
CNTs fails for graphene, especially in explaining the qualitative
difference between {\NOO} and the other adsorbates.

In this work we study the {\NOO} and {\NOOd}
adsorbate effects discussed above by combining \textit{ab-initio}
theory with transport measurements. Theoretically, the
electronic properties of carbon related materials and the
prediction of adsorbate geometries from first principles can be
well addressed by the density functional theory (DFT). Although
van der Waals forces are ill represented in the local density
approximation (LDA) as well as in gradient corrected exchange
correlation functionals (GGA) resulting in over- and underbonding,
respectively \cite{meijer:1996}, for polar molecules like {\NOO}
these errors are of minor importance: In similar studies of {\NOO}
on nanotubes both functionals yield qualitatively the same
predictions of adsorption energies and geometries \cite{YimW2003}.
Here, we apply both functionals, so that we obtain upper and lower
bounds for adsorption energies and related structural properties,
such as the equilibrium distances between molecules and graphene plane.

All DFT calculations were carried out with the Vieanna Ab Initio
Simulation Package (VASP) \cite{Kresse:PP_VASP} using projector
augmented waves (PAW) \cite{Kresse:PAW_VASP,Bloechl:PAW1994} for
describing the ion cores. The corresponding plane wave expansions
of the Kohn-Sham orbitals were cut off at $875$\,eV in the GGA
\cite{PW91_a, Perdew:PW91} and at $957$\,eV in the
LDA-calculations. In this periodic scheme, single {\NOO} and
{\NOOd} adsorbates are modelled in $3\times 3$ and $4\times 4$
graphene supercells, respectively. The ionic configurations
presented in this letter are fully relaxed, i.e. all forces being
less than $0.02$\,eV\,$\ang^{-1}$, and the convergence of
subsequent total energy calculations is guaranteed by applying the
tetrahedron method with Bl{\"o}chl corrections on
$\Gamma$-centered k-meshes denser than $30\times 30\times 1$, when
folded back to the single graphene Brillouin zone.

In the spirit of Ref. \cite{santucci:2003}, the DOS obtained in our DFT calculations are the central quantities
in the following discussion of the adsorbate effects on the
electronic properties of the graphene sheets. In particular, the
occurrence of a strong acceptor level due to \textit{individual}
{\NOO} molecules is predicted in this way, as well as
experimentally confirmed.

%Predictions for  are given and compared to new experiments on {\NOO} exposed graphene,in order to reveal especially the origin of the capability ofsingle molecule detection.

%\section{Theory Method}
%The electronic properties of graphene are studied by means of density functional theory (DFT) using the  and PW91 GGA exchange correlation functional. The single adsorbates on ideal graphene are modelled by $3\times 3$ and $4\times 4$ graphene supercells. For overlayers of {\HHO} and {\NHHH} a $(\sqrt{3}\times\sqrt{3})$\,R$30^\circ$ supercell is applied. The vacancy defects are studied using $5\times 5$ supercells with one atom missing. In contrast to all other adsorbates studied here, {\NOO} has one unpaired electron and will be investigated by spin-polarized DFT. As stopping condition for the ionic relaxations, all forces are required to be less than $0.02$\,eV\,$\ang^{-1}$. These fully relaxed structures are taken as input for the density of states (DOS), total energy and subsequent band structure calculations. The Brillouin zone integrations are performed on $\Gamma$-centred k-meshes denser than $15\times 15\times 1$ and $30\times 30\times 1$, when folded back to the single graphene Brillouin zone, for relaxations and DOS calculations, respectively. In that $50-100$\,meV Gaussian broadening and the tetrahedron method with Bl{\"o}chl corrections are applied, respectively.
\begin{figure}
\centering \resizebox{85mm}{!}{\includegraphics{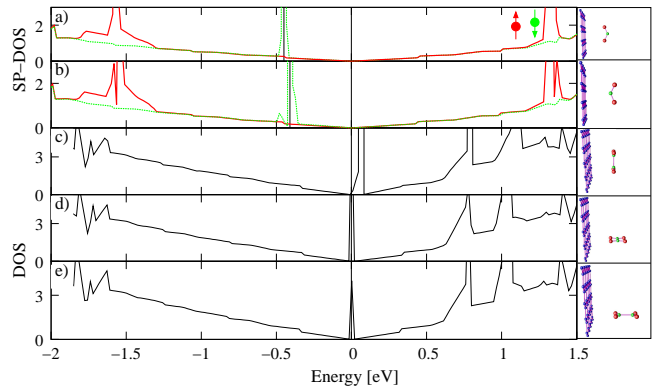}}
%\centering \resizebox{90mm}{!}{\includegraphics{DOS_NO2-perf3c.eps}}
\caption{\label{fig:NO2_adsorb_DOS} Left: Spin-polarised DOS of
the graphene supercells with adsorbed {\NOO}, a) - b),and DOS of
graphene with {\NOOd}, c) - e), in various adsorption geometries.
The energy of the Dirac points is defined as $E_D=0$. In the case
of {\NOO} the Fermi level $E_{\rm f}$ of the supercell is below
the Dirac point, directly at the energy of the spin down POMO,
whereas for {\NOOd} $E_{\rm f}$ is directly at the Dirac points.
Right: Adsorption geometries obtained with GGA. The carbon atoms
are printed in blue, nitrogen green and oxygen red.}
\end{figure}

%\section{DFT Results}
Gaseous {\NOO} stands in equilibrium with its dimer {\NOOd} giving
rise to various different adsorption mechanisms on graphene. For both, we
obtained possible adsorption geometries as depicted in Fig.
\ref{fig:NO2_adsorb_DOS} right. The corresponding adsorption
energies in GGA are $85$\,meV (a), $67$\,meV (b), $67$\,meV (c),
$50$\,meV (d) and $44$\,meV (e) per molecule with sheet-adsorbate
distances of $3.4-3.5\ang$ for the monomer and $3.8-3.9\ang$ for
the dimer. As usual, LDA yields higher adsorption energies - approximately
$169-181$\,meV for the monomer and $112-280$\,meV for the dimer -
and favours the adsorbates by $0.5-1\ang$ nearer to the sheet.

The spin-polarized DOS of the supercells
containing {\NOO},  shown in Fig. \ref{fig:NO2_adsorb_DOS} a) and
b), reveals a strong acceptor level at $0.4$\,eV below the Dirac
point due to these adsorbates in both adsorption geometries.

The molecular orbitals of {\NOO} correspond to flat bands and
manifest themselves as peaks in the DOS. The energies of these
peaks are virtually independent of the adsorbate orientation. Most
important for doping effects is the partially occupied molecular
orbital (POMO) of {\NOO}, which is split by a Hund like exchange
interaction: The spin-up component of this orbital is
approximately $1.5$\,eV below the Dirac point and fully occupied,
as it is also for the case of free {\NOO} molecule. The spin down
component of the {\NOO} POMO is \textit{unoccupied} for free
{\NOO}, but $0.4$\,eV below the Dirac point in the adsorbed
configuration \footnote{The energies of the POMO spin up and down
orbitals relative to the Dirac points predicted by LDA are
$-0.5$\,eV and $-1.4$\,eV, respectively, i.e. in almost
quantitative agreement with the GGA results.}. Hence, it can accept
one electron from graphene.

In contrast to the paramagnetic monomer, the dimer, {\NOOd}, has
no unpaired electrons and is diamagnetic: on formation from two
monomers the two POMOs hybridize. The resulting bonding orbital is
the highest occupied molecular orbital (HOMO), whereas the
antibonding combination forms the lowest unoccupied molecular
orbital (LUMO) for free {\NOOd}. The possibility of doping effects
due to adsorbed dimers has been investigated using the DOS depicted
in Fig. \ref{fig:NO2_adsorb_DOS} c) - e).

%\begin{figure}[ht]
%\centering \resizebox{110mm}{!}{\includegraphics{DOS_N2O4.eps}}
%\caption{\label{fig:DOS_bands_NO2d_adsorb_perf} DOS of supercells
%for the adsorbate geometries shown in Fig.
%\ref{fig:Struct_NO2_adsorb_perf}.c-e.}
%\end{figure}

Again, the molecular orbitals of the adsorbates are recognizable
as sharp peaks in the supercell DOS. One finds that the HOMO is in
all cases more than $3$\,eV below the Fermi level and therefore
does not give rise to any doping. However, the LUMO is always
quite near to the Dirac point, i.e. between $1$\,meV  and
$66$\,meV above it \footnote{LDA locates these LUMOs between
$2$\,meV and $210$\,meV above the Dirac point.}.

These initially empty orbitals can be populated by the graphene
electrons due to thermal excitations and act consequently as
acceptor levels. Thus \textit{both}, {\NOOd} and {\NOO}, give rise
to $p$-type doping of graphene ---  with one decisive distinction:
The affinity of the paramagnetic monomer to accept electrons from
graphene is much stronger than for the dimer, manifesting itself
in the energies of the  acceptor levels being well below and near
the Dirac point, respectively. This confirms exactly our
estimations that magnetism is crucial for a strong doping effect.
Without spin polarization, the quasilocalized states are typically
formed in the pseudogap of the host DOS: The resonant energy of an
impurity state is approximately located, where the real part of
the denominator of the scattering $T$-matrix vanishes
\cite{balatsky:373}. Here, however, it is Hund exchange splitting
which creates acceptor (or donor) peaks at distances of order of
half of an eV which is optimal for strong doping, as we observed
for the case of {\NOO}.

Experimentally, the existence of these two distinct types of
impurity levels was confirmed by combining electric field effect
and Hall measurements at different adsorbate concentrations. With
the experimental setup described in \cite{schedin-gassensors} we
measured the Hall resistance $R_{\rm xy}$ of an {\NOO}-exposed
graphene sample as a function of the gate voltage $V_{\rm G}$,
where different adsorbate concentrations are achieved by repeated
annealing of the sample at $410$\,K.

\begin{figure}
\centering
% \begin{tabular}{cc}
     \resizebox{80mm}{!}{\includegraphics{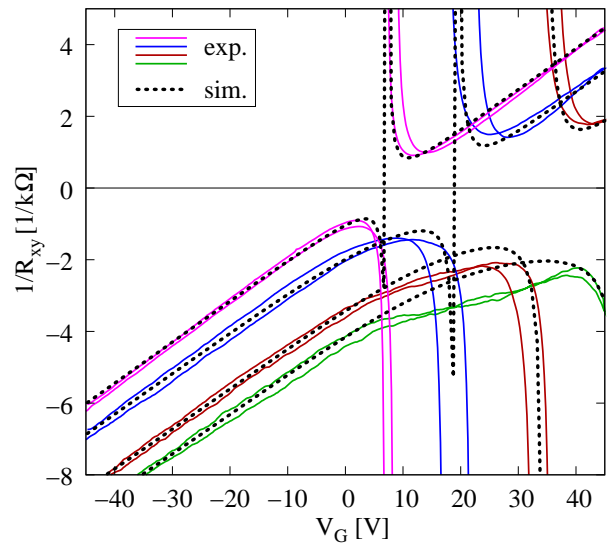}}
%     \resizebox{90mm}{!}{\includegraphics{Annealing.eps}}
% \end{tabular}
\caption{\label{fig:1Rxy_VG_NO2_adsorb} The inverse Hall resistance
$1/R_{\rm xy}$ is shown as a function of the gate voltage $V_{\rm G}$ for a
graphene sample with different concentrations of {\NOO} on top. The solid lines are the experimental
results with the margenta curve corresponding to the highest and the green curve to the
lowest concentration of adsorbates. The dotted lines are the simulations.  }% Right: {\NOO} and {\NOOd} concentrations on the sheet obtained by fitting the measurements to the simulations.}
\end{figure}

At magnetic fields $B=1$\,T and room temperature $T=290$\,K we
obtained $1/R_{\rm xy}$ versus $V_{\rm G}$ as depicted in Fig.
\ref{fig:1Rxy_VG_NO2_adsorb}. These measurements exhibit two
characteristic features.

\begin{figure}
\centering
\resizebox{80mm}{!}{\includegraphics{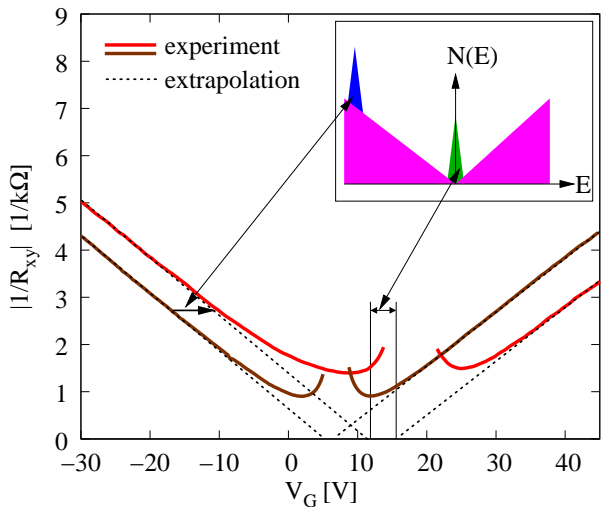}}
\caption{\label{fig:DOS_adsorbates_schema} The inset shows the DOS of graphene (magenta) with the strong acceptor level (blue) and the weak acceptor level (green) near the Dirac point. These impurity levels manifest themselves in the experimental
$1/R_{\rm xy}$ versus $V_{\rm G}$ curves shown in the main panel. After long annealing (84h) one
obtains the brown curve, whereas red curve corresponds to a higher adsorbate concentration on the sheet. The deep acceptor level causes a solid shift at all $V_G$, while the acceptor level close to the Dirac
point gives rise to an additional shift
only at larger positive gate voltages.}
\end{figure}

For gate voltages below $-10$\,V the curves are shifted in
parallel towards lower $1/R_{\rm xy}$ values when increasing the
absorbate concentration. Furthermore, the peak corresponding to
$R_{\rm xy}\rightarrow 0$ is shifted to higher gate voltages,
broadens and the minimal $|1/R_{\rm xy}|$ values increase with
more adsorbates on the sheet. We argue that this behaviour, as a
whole, is a direct evidence of the two distinct acceptor levels
(See Fig. \ref{fig:DOS_adsorbates_schema}.).

Depending on the adsorbate concentration and the gate voltage
there are in general three different types of charge carriers
contributing to the Hall effect: electrons and holes in graphene
as well as electrons in the impurity states. Assuming an equal
electron and hole mobility in graphene, the inverse Hall
resistance is given by $1/R_{\rm xy}=e\frac{(\mu
c+n+p)^2}{B(\mu^2c+n-p)}$, where $n$ ($p$) is the density of
electrons (holes) in graphene, $c$ is the density of electrons in
impurity states and $\mu$ their mobility in units of the graphene
electron mobility.

We determined $\mu\approx 0.08$ from our electric field effect measurements and the only adjustable is parameter
density of impurity states $N_{\rm imp}(E)$ yielding $c$ via the
Fermi distribution. As the gate voltage $V_{\rm G}=\alpha\sigma$
is directly related to the total charge density of the sample
$\sigma=e(c+n-p)$, where the prefactor $\alpha$ is determined by
substrate properties as described in \cite{Novoselov_science2004},
we can simulate the Hall resistance as a function of $V_{\rm G}$
by assuming an explicit form of the impurity DOS.

\begin{figure}
\centering
\resizebox{80mm}{!}{\includegraphics{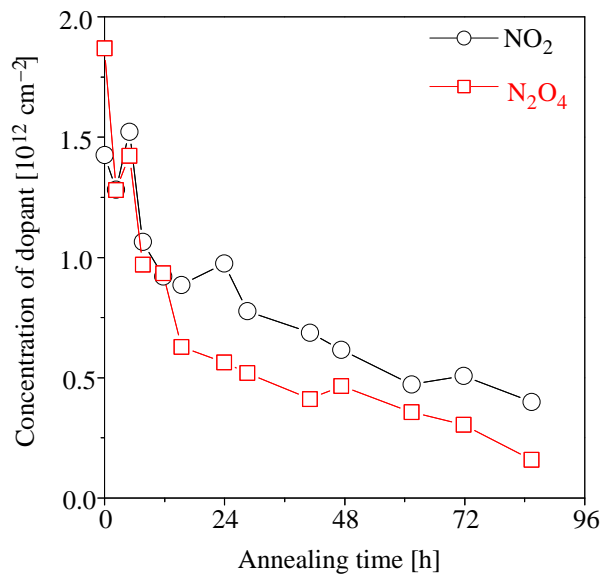}}
\caption{\label{fig:Charge_Anneal}
The measured adsorbate concentrations are shown as a function of annealing time. Those were obtained by identifying the {\NOO} ({\NOOd}) concentrations with the density of impurity states $c_1$ ($c_2$). The latter have been determined by fitting the simulated $1/R_{\rm xy}$ versus $V_G$ curves to our experiments.}
\end{figure}

It turns out, that for reasonable agreement of \textit{all} experimental curves
with the simulations $N_{\rm imp}(E)$ has to be peaked around
\textit{two} distinct energies, $E_1\lesssim -300$\,meV and
$E_2\approx -60$\,meV. Therefore, the experiment yields two types of acceptor levels, one rather close to the Dirac point and the other well below, in agreement with the DFT predictions. The simulations shown in Fig. \ref{fig:1Rxy_VG_NO2_adsorb} are
according to these impurity level energies, in particular $N_{\rm
imp}(E)=c_1\delta(E-E_1) + c_2\delta(E-E_2)$ \footnote{The broadness
of the impurity levels does not strongly influence the
simulations.}. Fitting the number of impurity states $c_1$ ($c_2$) at
$-300$\,meV ($-60$\,meV) to each experimental curve
yields the values for $c_1$ and $c_2$ depicted in Fig. \ref{fig:Charge_Anneal}.

One sees that the adsorption of monomers and dimers is similarly likely, so that understanding the interplay of adsorbate and gate voltage effects on the transport properties of {\NOO} exposed graphene requires accounting for monomers as well as dimers. Note also, that the similar adsorption and annealing rates for {\NOO} and {\NOOd} in experiment correspond nicely to the fact, that in both, GGA and LDA, the adsorption energies of the monomer are quite close to those of the dimer.

In \textit{conclusion}, we have elucidated the microscopic origin
of doping effects due to molecular adsorbates on graphene and work out the crucial relation between magnetic moments and doping strength in this context. In particular, we prove the occurrence of a strong acceptor level due to single
{\NOO} molecules which is associated with the formation of local
magnetic moment. Especially this strong acceptor level is the
origin of many interesting phenomena and a promising candidate for
tailoring the electronic and magnetic properties of future
graphene devices. Besides explaining the effect of single molecule
detection reported in \cite{schedin-gassensors} it is capable of
controlling the occupancy of flat impurity bands in graphene near
the Dirac point and can give rise to exchange scattering. This earns future attention, as it can result in strongly
spin-polarized impurity states \cite{wehling-2006-} and guides a possible pathway \cite{Edwards_2006}
to high temperature magnetic order in this material.
\bibliography{ref_ads.bib}
\end{document}